# Crystal structure prediction using evolutionary algorithms: principles and applications

Artem R. Oganov and Colin W. Glass

Laboratory of Crystallography, Department of Materials, ETH Zurich, Wolfgang-Pauli-Strasse 10, Zurich 8093, Switzerland

We have developed an efficient and reliable methodology for crystal structure prediction, merging *ab initio* total-energy calculations and a specifically devised evolutionary algorithm. This method allows one to predict the most stable crystal structure and a number of low-energy metastable structures for a given compound at any *P-T* conditions without requiring any experimental input. Extremely high (nearly 100%) success rate has been observed in a few tens of tests done so far, including ionic, covalent, metallic, and molecular structures with up to 20 atoms in the unit cell. We have been able to resolve some important problems in high-pressure crystallography and report a number of new high-pressure crystal structures (stable phases: ε-oxygen, new phase of sulphur; new metastable phases of carbon, sulphur and nitrogen, stable and metastable phases of $CaCO_3$). Physical reasons for the success of this methodology are discussed.

# I. Introduction.

Atomic structure is the most important piece of information about crystalline solids: just from the knowledge of topology of the structure, a precise structural model and many physical properties of crystals can be calculated with state-of-the-art quantum-mechanical methods. At the same time, prediction of the likely structural topologies (structure types) on fully theoretical grounds remains an unsolved problem. While in many cases it is possible to solve crystal structure from experimental data, theoretical structure prediction is crucially important for several reasons:

1) When experimental data are of poor quality for structure solution (defective or small samples, especially at high pressures and temperatures) theory provides the last resort;

2) Theory is the only way of investigating matter at conditions that cannot be studied with today's experimental techniques, e.g. at ultrahigh pressures;

3) The ability to predict crystal structures will open up new ways of materials design.

In the words of John Maddox [1]:

"One of the continuing scandals in the physical sciences is that it remains in general impossible to predict the structure of even the simplest crystalline solids from a knowledge of their chemical composition… Solids such as crystalline water (ice) are still thought to lie beyond mortals' ken".

These words still remain largely true, as evidenced by poor results of the latest blind test for crystal structure prediction[2]. The search for the stable structure (i.e. the structure corresponding to the global minimum of the free energy surface) is made difficult by the fact that the free energy surface is exceedingly multidimensional (the number of degrees of freedom is $3N+3$, where $N$ is the number of atoms in the unit cell) and has an overwhelming number of local minima separated by high barriers.

To get some feeling of the number of possible structures, let us consider a simplified case of a fixed cubic cell with volume $V$, within which one has to position $N$ identical atoms. For further simplification let us assume that atoms can only take discrete positions on the nodes of a grid with resolution $\delta$. This discretisation makes the number $C$ of combinations of atomic coordinates finite:

$$C = \frac{1}{(V/\delta^3)} \frac{(V/\delta^3)!}{[(V/\delta^3)-N]!N!} \qquad (1)$$

If $\delta$ is chosen to be a significant fraction of the characteristic bond length (e.g., $\delta = 1$ Å), $C$ would give an estimate of the number of local minima of the free energy. If there is more than one type of atoms, the number of different structures significantly increases. Assuming a typical atomic volume ~10 Å$^3$, and taking into account Stirling's formula ($N! \approx (\frac{N}{e})^n \sqrt{2\pi N}$), the number of possible structures for an element A (compound AB) is $10^{11}$ ($10^{14}$) for a system with 10 atoms in the unit cell, $10^{25}$ ($10^{30}$) for a system with 20 atoms in the cell, and $10^{39}$ ($10^{47}$) for the case of 30 atoms in the unit cell. One can see that these numbers are enormous and practically impossible to deal with even for small systems and increase exponentially with $N$. It is clear that point-by-point exploration of the free energy surface going through all possible structures is not viable, except for the simplest systems with ~1-5 atoms in the unit cell.

Fortunately, one does not have to explore the entire free energy surface in order to locate the global minimum – it would suffice to explore just the most promising regions of that surface, and there are a number of methods for doing this. One either has to start already in a good region of configuration space (so that no effort is wasted on sampling poor regions), or use a "self-improving" method that locates, step by step, the best structures. The first group of methods includes metadynamics[3,4], simulated annealing[5,6], basin hopping[7] and minima hopping[8] approaches. The second group essentially includes only evolutionary algorithms[9-12]. While both groups of methods have their advantages and disadvantages, our personal view is that evolutionary algorithms are more suitable for the task. The strength of evolutionary simulations is that they do not necessarily require any system-specific knowledge (except chemical composition) and are self-improving, i.e. in subsequent generations increasingly good structures are found and used to generate new structures. This allows a "zooming in" on promising regions of configuration space. Flexible nature of the variation operators allows one to incorporate features of other methods into an evolutionary algorithm.

Section II outlines features desirable for a successful method; most of these features are present in our proposed method. Aspects of free energy landscapes important for global optimisation are discussed in Section III and our method is described in Section IV. Section V presents tests of our method on well-studied systems and predictions of several previously unknown high-pressure crystal structures. Reasons of success of this method and future work are discussed in Section VI.

## II. Properties of the desired method.

With standard *ab initio* techniques, to predict the stable structure one usually considers structures known for similar systems[13] and finds the phase with the lowest free energy. When unexpected or previously unknown structures become stable, this approach fails – literature abounds with examples of such failures[20].

We want to develop a systematic approach, where the stable crystal structure at given *P-T* conditions is found using only the chemical composition. This method should be universal – applicable to ionic, metallic, covalent or van der Waals crystals. Different types of structures (e.g. molecular, chain, layered, framework) should be treatable on an equal footing.

It is desirable to be able to predict also low-energy metastable structures[21]. In this case, the knowledge of phase transition mechanisms is useful for guiding synthesis. While idealised phase transition mechanisms can be simulated with such methods as metadynamics (see e.g. Ref. 22), real mechanisms are much more complicated and usually involve nucleation and growth, which are difficult to treat.

The method should scale well with $N$ – although an exponential scaling is almost inevitable at large $N$. Many interesting high-pressure crystal structures contain up to ~40 atoms in the unit cell – thus, the method that we need should be efficient at least up to $N \approx 40$. Sometimes (especially for organic crystals and complex metallic alloys) unit cells contain hundreds or thousands of atoms, and good performance for such structures is valuable.

The method should have a reliable behaviour – up to a certain system size a realistic calculation should yield a sufficiently high success rate, allowing verification by experiment or a second run. A more controversial criterion is the guaranteed convergence behaviour – with an infinite-size calculation one should always find the correct structure. Increasing the size of the calculation would increase confidence in the results and provide a way to check them.

Finally, it is important that the method be easily parallelisable. Ideally, operations on different processors should be completely independent of each other, and the workload equally distributed.

## III. Search space and energy landscape.

Free energy landscapes in chemical systems have an overwhelming number of local minima separated by high barriers, which precludes molecular dynamics or Monte Carlo approaches from being generally useful for crystal structure prediction. However, these landscapes have several benign general properties:

1. Only a tiny fraction of the landscape is of any chemical interest (i.e. corresponds to reasonable bond lengths etc.).

2. One structure in general corresponds to many equivalent minima of the free energy (due to crystal symmetry and the non-uniqueness of the choice of the unit cell).

3. The deepest minimum has the greatest surface area and is thus easier to reach than any single local minimum (however, the number of different local minima is overwhelmingly large).

4. Most or all of the low-energy minima are located in the same region of the landscape. This gives an overall shape to the landscape and stems from the fact that most good structures have similar bond lengths and coordination environments of the atoms, i.e. are made of similar building blocks.

5. Low-lying free energy minima are usually separated by low energy barriers (Bell-Evans-Polanyi principle).

Most methods designed so far for crystal structure prediction – simulated annealing, metadynamics, minima hopping and basin hopping, have focussed on overcoming the (free) energy barriers. Usually, in these methods one starts already in a good region of configuration space and explores its neighbourhood by moving across a barrier into a nearby local minimum, hoping to discover a new good minimum (*cf.* landscape property 5). The problem of global search through a huge number of local minima is replaced here by a neighbourhood search through a relatively small number of local minima. This may bring computational advantages, but will only work efficiently (or at all for some methods) when the stable structure is very similar to the starting structure. Besides, a good starting structure is not always *a priori* known. Metadynamics[3,4] is a powerful method that reduces the dimensionality of the problem to a small number of relevant order parameters – most commonly, lattice vectors are used as a 6-dimensional order parameter. This method is often successful and efficient, but has known failures and the choice of an order parameter is not always clear. Another neighbourhood search method, simulated annealing[5,6], has a guarantee of finding the global minimum in an infinite calculation – however, this method involves intricate parameter tuning, is computationally very expensive and has little potential in the framework of *ab initio* calculations.

Evolutionary algorithms (e.g., Ref. 23,24) present an attractive alternative to these methods, since for certain types of problems they are known to be very efficient at finding global minima of multidimensional functions, are fully non-local, do not involve any assumptions on the order parameter or topology of the landscape, and no initial structure is required. These algorithms mimic Darwinian evolution and employ natural selection of the fittest and such variation operators as genetic heredity[25] and mutations. Evolutionary algorithms can perform well for different types of free energy landscapes, but are especially powerful for landscapes having an overall shape – those where the global minimum and the deepest local minima are located in the same region of configuration space, which is expected for chemical systems (landscape property 4). Evolutionary algorithms have been applied to crystal structure prediction[9-11] for the case of fixed lattice parameters, employing a cheap heuristic evaluation function, a discrete grid for atomic positions, and encoding structural information by binary "0/1" strings. In spite of many simplifications, this approach turned out to be computationally expensive. While Bush *et al.*[9] succeeded in solving the structure of $Li_3RuO_4$ using this method and experimental lattice parameters, extensive tests[10,11] showed that it often fails even for simple structures –e.g. $TiO_2$ anatase[11].

We have developed a new, highly efficient and reliable, evolutionary algorithm which does not require any experimental information. This method is reminiscent in some features of the method used by Deaven & Ho[12] to predict structures of molecules[26] and incorporates some ideas from Ref. 3,4,11 and a large number of our own developments. This combination of ideas has resulted in a very powerful method that satisfies many criteria discussed in the previous section: (i) it achieves reliable structure prediction from the knowledge of the chemical composition alone, (ii) predicts not only the stable structure, but also a set of low-energy metastable phases, (iii) is efficient and is easily combined with *ab initio* calculations, and (iv) possesses excellent parallelisability. There is a guarantee that an infinitely big calculation will locate the global minimum. However, our method provides no information on mechanisms of phase transitions – to study these one has to use other methods.

## IV. Our implementation: the USPEX code.

Our algorithm has been implemented in the USPEX (Universal Structure Predictor: Evolutionary Xtallography) code – for a detailed account see Ref. 27. The code has a minimal input: (i) the number of atoms of each sort, (ii) pressure-temperature conditions and such algorithm parameter values as (iii) the size of the population (i.e. the number of structures in each generation), (iv) hard constraints[28], (v) the number of structures used for producing the next generation, and (vi) the percentage of structures obtained by lattice mutation, atomic permutation and heredity. Other parameters define how often structure slices combined in heredity are randomly shifted, how many atomic permutations are done on average per structure, and strength of lattice mutation. Optionally, calculations can be performed under fixed lattice parameters (if these are known from experiment).

USPEX represents lattice vectors and atomic coordinates by real numbers (not by the often used binary strings), which enhances the learning power of the algorithm. Every candidate structure is locally optimised and subsequently replaced by the locally optimal structure. For local optimisation we use conjugate-gradients or steepest-descent methods, available in many first-principles and atomistic simulation codes. Currently, USPEX can use VASP[29], and SIESTA[30] for first-principles and GULP[31] for atomistic simulations. During *ab initio* optimisation the k-points grid changes in accordance with cell changes, enhancing comparability of the free energies.

Our procedure is the following:

1. The first generation is produced randomly (but only structures satisfying the hard constraints are allowed). Starting from user-provided structures is also possible.

2. Among the locally optimised structures, a certain number of the worst ones are rejected, and the remaining structures participate in creating the next generation through heredity, atomic permutation and mutation. Selection probabilities for variation operators are derived from the fitness ranking of the structures (i.e. their free energies).

During heredity, new structures are produced by matching slices (chosen in random directions and with random positions) of the parent structures. Some of the structures are produced by randomly shifting slices in their matching plane. Heredity for the lattice vectors matrix (represented in the upper-triangular form to avoid unphysical whole-cell rotations) is done by taking a weighted average, with a random weight.

A certain fraction of the new generation is created by permutation (i.e. switching identities of two or more atoms in a structure) and mutation (random change of the cell vectors and/or atomic positions). For atomic permutation the number of permutations applied to a given structure is drawn from a uniform distribution. For lattice mutation we define each mutated cell vector **a'** as a product of the old vector ($\mathbf{a}^0$) and the ($\mathbf{I}+\boldsymbol{\varepsilon}$) matrix:

$$\mathbf{a'} = (\mathbf{I}+\boldsymbol{\varepsilon})\mathbf{a}^0 , \qquad (1)$$

where **I** is the unit matrix and $\boldsymbol{\varepsilon}$ is the symmetric strain matrix, so that:

$$(\mathbf{I}+\boldsymbol{\varepsilon}) = \begin{pmatrix} 1+\varepsilon_1 & \varepsilon_6/2 & \varepsilon_5/2 \\ \varepsilon_6/2 & 1+\varepsilon_2 & \varepsilon_4/2 \\ \varepsilon_5/2 & \varepsilon_4/2 & 1+\varepsilon_3 \end{pmatrix} \qquad (2)$$

The strain matrix components (and the values of atomic position shifts) are selected randomly from the Gaussian distribution and are only allowed to take values between -1 and 1. Lattice mutation essentially incorporates the ideas of metadynamics[3,4] into our method. Metadynamics finds new structures by building up cell distortions of some known structure, but in our method distortions are not accumulated and to yield new structures the strain components should be large.

To avoid pathological lattices all newly obtained structures are rescaled to have a certain volume, which is then relaxed by local optimisation. The value of the rescaling volume can be obtained from the equation of state of some known structure or by optimising a random structure; this value is used only in the first generation and for subsequent generations is adapted to the volumes of several best found structures. A specified number of the best structures (usually, one) in the current generation survive into the next generation.

3. The simulation is stopped after some halting criterion is met. In our experience, for systems with ~20 atoms in the cell finding the stable crystal structure usually takes up to ~20 generations (Fig. 1).

Among the important results of the simulation are the stable structure and a set of low-energy structures at given pressure-temperature conditions. In our experience, this method is efficient even in such problematic situations as finding correct atomic ordering (thanks to atomic permutations) and polytypism (thanks to spatial heredity building new structures from slabs of parent structures). Though it is possible to calculate finite-temperature free energies also from first principles[32] and thus use USPEX at finite temperatures, *ab initio* simulations reported here are done at 0 K (which makes simulations much more affordable).

## V. Tests and applications.

Here we discuss tests of USPEX on systems with well-known stable structure, and some applications that have revealed hitherto unknown structures. All the calculations discussed below were performed in the framework of density functional theory within the generalised gradient approximation[34] and the all-electron PAW method[35,36] as implemented in the VASP code[29] (see footnote[37]).

**Carbon at 0-2000 GPa.** Elemental carbon was mentioned by J. Maddox[1] as a challenging case for crystal structure prediction. Predicting the stability of the graphite structure at 0 GPa can be problematic, because of the incomplete description of the van der Waals forces (that binding graphite layers together) in *ab initio* simulations. The best structure found by USPEX is indeed built of flat graphite layers with a slightly displaced stacking and the same (within the numerical uncertainties of our calculations) total energy as graphite with the ideal stacking of the layers. Simulations at high pressures (100, 300, 500, 1000, 2000 GPa) did not show graphite (even among the metastable structures) and yielded other stable structures – diamond below 1000 GPa and "bc8" structure above 1000 GPa. This phase transition places an upper limit of 1000 GPa to the applicability of the diamond anvil cell technique in high-pressure experiments. The bc8 structure was known for a long time[38] as a metastable structure of Si, and was believed to be stable above 1000 GPa [39] – our simulations strongly support this idea. USPEX does not find the R8 structure, earlier claimed to be stable above 500 GPa [40]; we investigated this issue and found that the R8 structure is never stable at 0 K.

Simulations at 1000 GPa and 2000 GPa show very diverse metastable structures – tetrahedral structures such as diamond and higher-coordinated (e.g., simple cubic) structures. In the simulation at 1000 GPa (almost exactly the conditions of the diamond-bc8 transition) diamond and bc8 structures survived and competed all the way to the end of the run. At 100 GPa we see lonsdaleite ("hexagonal diamond") among metastable structures, this metastable polytype of diamond is experimentally known at high pressures. Another low-energy metastable structure found at 100 GPa and containing 5- and 7-fold rings is shown in Fig. 2; interestingly, this structure corresponds to the (2x1)-reconstructed of the (111) surfaces of diamond and silicon. All these simulations were performed with 8 atoms in the unit cell, and no information on cell dimensions or shape was used.

Exploring results at 0 GPa (4 and 8 atoms/cell), we found an extremely rich chemistry with competitive structures having tetrahedral and/or triangular coordination of carbon atoms. Among these, we found tetrahedral ($sp^3$-hybridization) structures with simultaneously present 4- and 8-membered rings, $sp^2$-structure with 8- and 4-membered rings forming a flat layer, $sp^2$-structures with 5- and 7-fold rings (resembling the structure type of $ScB_2C_2$) and an energetically reasonable structure consisting of linear carbyne chains (sp-hybridized). Two other interesting structures contained walls of the graphite structure interconnected by layers of the diamond structure or $sp^2$-carbon atoms in a different orientation (Fig. 3).

**Sulphur at 12 GPa.** Phases of sulphur known at ambient pressure, α-S and β-S, consist of nearly perfect crowned ring $S_8$ molecules. It is clear that such molecular structures, having large empty space inside the molecular rings, have low densities and will transform to denser phases at relatively low pressures. Indeed, above 2.5 GPa α-S (space group *Fddd*) transforms into the S-II structure consisting of spiral trigonal chains[41]. The structural sequence on increasing pressure has long remained controversial, largely due to metastability problems: in early works[42], some 12 phases have been seen in the small pressure range 0-4 GPa. More recent works[41,43-46] have found only the following high-pressure phases to be stable: trigonal S-II (space group $P3_221$) and rhombohedral ($R\bar{3}$) phase with $S_6$ molecules in the range from 3 GPa to 20-40 GPa, tetragonal S-III (space group $I4_1/acd$) phase stable up to ~95 GPa, and incommensurate monoclinic S-IV phase above 95 GPa. However, the phase diagram (especially its low-temperature part) remains poorly constrained. Compressed at room temperature, for kinetic reasons α-S transforms directly to S-III at 38 GPa – i.e. skipping phases S-II and $R\bar{3}$.

To gain better insight into structural stability of sulphur, we have explored its behaviour at 12 GPa using USPEX. Given that the triple point between phases S-II, $R\bar{3}$ and S-III was located[46] at similar conditions (10.5 GPa and 650 K), very different structures are expected to be energetically competitive at this pressure – thus posing a stringent test to our method. Simulations with 3, 4, 6, 8, 9, 12 atoms/cell and using no experimental input were performed and resulted in ~20 energetically similar (yet structurally very different) phases. The structures found include all the structures experimentally observed between ~3-90 GPa ($R\bar{3}$, Fig. 4a; S-II, Fig. 4d; S-III, Fig. 4b), Se-I-type structure (Fig. 4e) and some entirely new structures (e.g. Fig. 4c,f). One of the new structures (Fig. 4c) emerges from our calculations as thermodynamically stable at 0 K and pressures up to 7 GPa – though more favourable than S-II by only 5 meV/atom. The main elements of these structures (trigonal and tetragonal spiral chains and $S_6$ crowned ring molecule) are shown in Fig. 5, and their energetics are compared in Fig. 6. At 7-18 GPa the $R\bar{3}$ structure is stable at 0 K, and above 18 GPa the most stable phase is tetragonal S-III.

These results confirm the power of our method and suggest that more experiments should be done to explore the phase diagram of sulphur and the possible stability field of the twisted-chain structure of Fig. 4c. Its structure at 12 GPa is: space group $P2_12_12_1$, $a$ = 6.862 Å, $b$ = 4.684 Å, $c$ = 4.375 Å, S1=(0.0945;0.1163;0.0744), S2=(0.2974;0.2474;0.4332).

**Other elements under pressure: hydrogen, nitrogen, oxygen, iron, xenon.** Metallisation of hydrogen under pressure has attracted much attention since the prediction of Wigner and Huntington[47] that it will occur at pressures not lower than 25 GPa. With experiments having explored pressures above 300 GPa [48,49], solid metallic hydrogen still has not been observed. The most popular picture of the high-pressure behaviour of solid hydrogen includes insulating molecular phases I-III, then *metallic molecular* solid at pressures above ~350 GPa and *metallic non-molecular* solid above ~500 GPa [50,51]. The structure of phase II has been recently proposed[52], but essentially nothing is known about the structures of phase III and higher-pressure metallic phases. From our preliminary results it emerges that even at the pressure of 600 GPa hydrogen keeps the molecular structure: all the generated structures at this pressure were molecular. While *molecular metallic* hydrogen is possible at such pressures, stable *non-molecular metallic* hydrogen requires pressures much higher than 600 GPa.

The low-pressure structures of nitrogen contain triply-bonded $N_2$ molecules with exceptionally high dissociation energy of 9.76 eV, but it is well understood that under pressure molecular structures must transform into non-molecular ones[53]. Based on stereochemical considerations and some imagination, Mailhiot and co-workers[54] came up with an ingenious candidate non-molecular structure for nitrogen, which they christened "cubic gauche". Mailhiot *et al.*[54] predicted that this structure will become more stable than known molecular structures above 50 GPa. After many experimental attempts this phase has eventually been synthesised[55], but since experimental data on the high-pressure phase diagram of nitrogen remain unclear or even controversial[56], we have decided to investigate the stability of the *cubic gauche* structure using USPEX. We ran calculations at 100 GPa with 8 atoms in the unit cell. The most stable structure found in this run was indeed *cubic gauche*, and we also found a number of low-energy metastable chain (previously obtained in Ref. 57) and layered structures, as well as another 3D-polymeric structure related to *cubic gauche* (Fig. 7) and less stable by 49 meV/atom.

For oxygen, a number of high-pressure phases are known at pressures up to ~150 GPa [58,59]; all of these structures are molecular. Structures of the red ε-phase (8-96 GPa) and black metallic (at low temperatures superconducting[60]) ξ-phase (above 96 GPa) are still uncertain, in spite of numerous experimental measurements (X-ray and neutron diffraction, vibrational and electronic spectra). Latest experiments indicate that magnetism of oxygen disappears with the formation of the ε-phase[61], in agreement with an earlier theoretical study[62] which found a non-magnetic ground-state structure. However, a different structure with paired $O_2$ molecules forming $O_4$ units was deduced from vibrational spectra of ε-phase[63]. We performed a number of simulations with different numbers of atoms in the unit cell (up to 16) and at different pressures (25, 130, 250 GPa) and found the structures of the ε- and ξ-phases, as well as a new stable crystal structure in the run at 250 GPa. Surprisingly, even at 250 GPa the structure is still molecular.

Our most stable structure is the same as the one reported in Ref. 62. This structure (Fig. 8a) consists of zigzag chains made of $O_2$ molecules. The intermolecular O-O distance is 2.04 Å – much longer than the intramolecular O-O bond length of 1.22 Å. At 25 GPa we also found a very competitive metastable structure containing $O_4$ groups like the structure of Gorelli *et al.*[63]. Both structures have IR- and Raman-active modes, but the $O_4$-structure produces X-ray diffraction patterns much more similar to experimental[59]. We conclude that the phase that has been seen experimentally[58,59] contains $O_4$ units in agreement with Ref. 63, but seems to be metastable (which is possible since this phase was obtained in experiments at room temperature). Both structures are unique and not found for any other element so far; both show a large difference between inter- and intramolecular O-O distances and must be considered as molecular.

For Fe at pressures of the Earth's inner core (~350 GPa) and 0 K we predict the hcp-structure to be stable, in agreement with most recent studies[64]. At high temperatures, a transition to the bcc- structure may be possible[65,66] – to investigate this possibility, finite-temperature USPEX simulations will be needed.

For Xe the most stable structure found at 200 and 1000 GPa is also the hcp-structure, in agreement with experiments. Experiments suggest that metallization of xenon[67] at 130 GPa preserves its hcp-structure[68] – our results support this picture.

**$MgSiO_3$ perovskite and post-perovskite.** As shown in Fig. 1, for $MgSiO_3$ we correctly predict the $CaIrO_3$-type post-perovskite structure at 120 GPa: according to *ab initio* simulations[16] and experiments[19,16] this structure is more stable than perovskite above 100 GPa (USPEX simulations still find it as the most stable structure at 1000 GPa). Simulation at 120 GPa also identifies the perovskite phase, metastable at this pressure. At 80 GPa USPEX correctly produced the perovskite structure as stable. Interestingly, most configurations found at 120 GPa contained chains of edge-sharing $SiO_6$-octahedra (like in the post-perovskite structure), which indicates that such structures become favourable at very high pressures. At 80 GPa, most structures contained corner-sharing octahedra (as in perovskite), although edge sharing was also occasionally observed. The possibility to "scan" the chemistry of the system at given conditions by looking at the stable structures is an important feature of our method.

**Al$_2$O$_3$ at 300 GPa.** For Al$_2$O$_3$, experiment[17] and simulations[17,18] demonstrated that the CaIrO$_3$-type structure is stable above ~130 GPa. This pressure corresponds to the conditions where shock-wave experiments evidenced a sudden drop of the electrical resistivity[69] and a small increase of density[70]. Further (and much greater) density anomaly was observed at 250 GPa [70] and tentatively attributed to a solid-solid phase transition. At least for some materials, e.g., MgSiO$_3$ and NaMgF$_3$, the CaIrO$_3$-type structure seems to be the ultimate high-pressure structure before decomposition[33], but decomposition can be ruled out for chemically simple Al$_2$O$_3$. In search of a post-CaIrO$_3$ type phase, we performed variable-cell evolutionary simulations at 300 GPa, but the result was that the CaIrO$_3$-type structure is stable at these conditions (in the same simulation we also observed a much less stable perovskite structure). This leads us to suggest that the phase transition seen in shock-wave experiments at 250 GPa (and very high temperatures[71]) was melting, rather than any solid-solid transition.

**SiO$_2$ at 0 GPa.** For silica at 1 atm (simulations with 9 atoms/cell), our variable-cell simulations correctly find quartz as the most stable structure. As metastable structures, we find various quartz-like structures, less stable layered tetrahedral structures, and even less stable structures including one containing simultaneously SiO$_4$-tetrahedra and SiO$_6$-octahedra. All the found reasonable metastable structures (within 0.1-0.2 eV/atom of the global minimum, quartz) are based on corner-sharing SiO$_4$-tetrahedra, in accordance with Pauling's rules. Structures contradicting Pauling's rules (e.g. those containing edge-sharing tetrahedra or those with octahedral coordination at ambient pressure) come out to be energetically very unfavourable.

**TiO$_2$ anatase.** Previous attempts to use evolutionary algorithms within the fixed-cell implementation (see Ref. 11) found that it quite difficult to reproduce the structure of anatase (12 atoms/cell), the main problem being the choice of a crude evaluation function. With USPEX we easily find this structure (Fig. 9).

**CaCO$_3$ at 50-150 GPa.** Chronologically the first application of USPEX was the solution of a non-trivial and hitherto unknown structure of the post-aragonite phase of CaCO$_3$ (Ref. 72). Experimental studies found this phase to be stable above 40 GPa [73], but could not solve its structure. Post-aragonite is an important mineral phase, likely to be a major host of carbon in the Earth's lower mantle. As one can see in Fig. 10, the structure found by USPEX closely reproduces experimental measurements and in agreement with experiment[73] it was predicted to be more stable than aragonite above 42 GPa. The same structure matches diffraction patterns of SrCO$_3$ (Ref. 74) and BaCO$_3$ (S. Ono, pers. comm.) at high pressure. Above 137 GPa, our simulations predict stability of a new structure (space group $C222_1$, Fig. 9c) containing carbon in the tetrahedral coordination. This prediction has been verified (S. Ono, pers. comm.) several months after publication[72]. Some metastable phases identified for CaCO$_3$ are shown in Fig. 11.

**Molecular crystals.** It can be expected that the most challenging materials for this methodology would be molecular crystals. There are two reasons: first, the availability of large "empty" space (i.e. the chance of putting an atom by mistake in this "empty" space is high), second, the possibility of creating different molecules from the same set of atoms – e.g., in the case of urea, the correct (NH$_2$)$_2$CO molecules or a set of N$_2$, H$_2$, O$_2$, CO$_2$, NH$_3$ molecules etc. Often these different possibilities would be energetically competitive – in such cases one could expect that the overall shape of the free energy landscape would contain several nearly equal attractor basins and the simulation could end up in a basin not containing the global minimum. However, we find that the method works well for CaCO$_3$ containing molecular carbonate-ions and for simple molecular solids (we have done simulations on hydrogen, oxygen, nitrogen, chlorine, fluorine). An interesting test case is ice, H$_2$O, which at atmospheric pressure adopts the "ice Ih" structure (where the oxygen sublattice is isostructural with lonsdaleite polymorph of carbon – a cubic diamond-type "ice

Ic" form is also known experimentally). Together with graphite, prediction of this structure is, according to J. Maddox[1], "still thought to lie beyond the mortals' ken". One can expect some difficulties for this material – low density and relatively weak bonding between the molecules make the problem quite challenging. Furthermore, the structure of ice is topologically frustrated, since it cannot simultaneously fulfil the "ice rules" and keep the full symmetry of the oxygen sublattice. As a result, the structure is disordered and characterised by partial occupancies of hydrogen positions. Disordered structures cannot be treated by USPEX – instead, the method produces the lowest-free energy ordered structure. Fixing the unit cell parameters at experimental values for a 12-atom cell, we correctly obtain the ice Ih structure with correctly fulfilled ice rules. Interestingly, with variable cell parameters for a 12-atom cell we find both the hexagonal (Ih) and cubic (Ic) structures of ice (Fig. 12), with energies within the uncertainties of the calculation. Both structures fulfil the ice rules and are proton-disordered.

For an even more challenging case of urea, $(NH_2)_2CO$ with 16 atoms in the unit cell, we easily find the correct structure when the unit cell parameters were fixed at experimental values. This success is far from trivial: during the simulation we observed a very large number of different molecules and with different orientations in the cell. These structures had similar energies, but the lowest-energy structure was (correctly) the tetragonal urea structure (Fig. 13). Without any experimental information, the simulation succeeds in identifying the correct molecules of urea[75], but it takes much longer to find their correct orientations in the unit cell. The energies of whole-molecule rotation are quite small compared to intramolecular bond energies, thus making the process of finding correct molecular orientations difficult. Fixing the molecular geometry (or at least bond connectivity) and using molecular rotations, translations and inversions as search parameters would greatly speed up structure predictions for molecular crystals.

## VI. Discussion.

This study shows that evolutionary algorithms are very powerful for crystal structure prediction. The evolutionary method that we have developed[27] allows one to find the stable crystal structure of a given compound at given external conditions (pressure, temperature, etc.), and provides a set of low-energy metastable structures. Unlike traditional simulation methods that only sample a small part of the free energy landscape close to some minimum, our method explores the entire free energy surface – on which it locates the most promising areas. This allows one to see which aspects of structures (molecular vs coordination or metallic vs insulating structures, atomic coordination numbers, bond lengths and angles) are required for stability and therefore provides an interesting way of probing structural chemistry of matter at different conditions. In its present implementation our method is very efficient for systems with up to ~30-40 atoms/cell, but with additional developments it should be possible to achieve efficiency also for larger systems.

The reasons why this method is so successful are rooted in the "good" properties of free energy landscapes mentioned in Section III. Random generation of the first population ensures unbiased sampling of the landscape, and thanks to local optimisation these structures are projected into the chemically interesting part of the landscape (property 1). Local optimisation is very effective for reducing the noise of the landscape. Relatively large basin area of the global minimum further increases the chance of sampling the global minimum and for systems with up to ~12 atoms/cell the global minimum is often reached already in the first few generations. Combining fragments of the selected optimised structures is likely to produce other reasonable structures. By selecting only the best structures (usually, the best ~60% of the structures are selected to produce the next generation) introduces an exponential

"zooming in" on the most promising area of the configuration space until the global minimum is found (property 4). Since no symmetry constraints are imposed during simulations, symmetry is one of the results of our algorithm. This ensures that the resulting structures are mechanically stable and do not contain any unstable Γ-phonons.

Our approach enables crystal structure prediction without any experimental input. Essentially, the only input is the chemical composition. However, one often needs to predict also possible stoichiometries –e.g. for metallic alloys, where it is very difficult even to rationalise stoichiometry. One of the first steps in this direction was taken in Ref. 76, where an *ab initio* evolutionary algorithm was used to find stable alloys with fcc- and bcc-structures. We plan to incorporate variable composition in our algorithm in order to find stable structures *and* likely stoichiometries at the same time. The present limitation of the method to ordered periodic structures can be overcome once it becomes possible to calculate free energies of disordered and aperiodic structures.

Clearly, the quality of the global minimum found by USPEX depends on the quality of the *ab initio* description of the system. Present-day DFT simulations (e.g., within the GGA) are adequate for most situations, but it is known that these simulations do not fully describe van der Waals bonding and the electronic structure of Mott insulators. In both areas there are significant achievements (see, e.g., Ref. 77-83), which can be used for calculating *ab initio* free energies and evaluation of structures in evolutionary simulations.

**Acknowledgements.**

We gratefully acknowledge funding from the Swiss National Science Foundation (grant 200021-111847/1) and access to computing facilities at ETH Zurich and CSCS (Manno). We thank N. Hansen and W.A. Crichton for useful discussions, T. Racic, G. Sigut and O. Byrde for computational assistance and M. Valle for developing the STM3 library[84] for the visualisation of our results.

spectra calculated from density-functional perturbation theory; for a review see [Baroni S., de Gironcoli S., Dal Corso A., Gianozzi P. (2001). *Rev. Mod. Phys.* 73, 515-562], for applications see Ref. 16,17,33 and [Oganov A.R., Gillan M.J., Price G.D. (2003). *J. Chem. Phys.* 118, 10174-10182; Oganov A.R., Gillan M.J., Price G.D. (2005). *Phys. Rev.* B71, 064104]. Another strategy is based on thermodynamic integration techniques; for a review see [Frenkel D., Smit B. (2002). *Understanding Molecular Simulation. From Algorithms to Applications.* San Diego, Academic Press], used e.g. in [Alfé D., Gillan M.J., & Price G.D. (1999). *Nature* 401, 462-464].

# Figure Captions.

**Figure 1. Prediction of the crystal structure of MgSiO$_3$ at 120 GPa (20 atoms/cell).** Enthalpy (per 20 atoms) of the best structure as a function of generation. Insets show the perovskite and post-perovskite structures (blue polyhedra – SiO$_6$ octahedra; gray spheres – Mg atoms). Between 6$^{th}$ and 12$^{th}$ generations the best structure is perovskite, but at the 13$^{th}$ generation the post-perovskite structure is found (at 120 GPa this structure is 0.3 eV lower in enthalpy than perovskite). This simulation used no experimental information and illustrates that our method can find both the stable and low-energy metastable structures in a single simulation.

**Figure 2. Structures of carbon identified by USPEX.** a) Graphite (0 GPa, stable state), b) Diamond (stable state up to 1000 GPa, metastable above 1000 GPa), c) bc8 structure (stable above 1000 GPa; structural parameters at 2000 GPa: space group *Ia*3, *a*=3.276 Å, $x_C=y_C=z_C$=0.1447), d) lonsdaleite (metastable at 100 GPa), e) "5+7" structure (metastable at 100 GPa, with structural parameters: space group *C*2/*m*, *a*=8.612, *b*=2.375, *c*=3.933 Å, $\beta$=96.59°; C1=(0.444,1/2,0.122), C2=(0.443,0,0.348), C3=(0.287,1/2,0.940), C4=(0.272,0,0.416).

**Figure 3. Metastable "wall" structures of carbon found by USPEX at 0 GPa.** a) structure containing both sp$^3$- and sp$^2$-hybridised carbon atoms (0.4 eV/atom higher in energy than graphite), b) structure containing only sp$^2$-carbon atoms (0.5 eV/atom higher in energy than graphite).

**Figure 4. Some of the energetically competitive high-pressure structures of sulphur found by USPEX.** a) $R\bar{3}$ phase with S$_6$ ring molecules, b) S-III phase with spiral tetragonal chains, c) $P2_12_12_1$ structure with distorted spiral tetragonal chains in twisted orientations, d) S-II phase with trigonal spiral chains in three orientations, e) Se-I structure with trigonal spiral chains in one orientation (space group $P3_121$), f) *C*2 structure with distorted spiral tetragonal chains running in two perpendicular directions.

**Figure 5. Structural elements of high-pressure phases of sulphur.** a) spiral trigonal chain, b) spiral tetragonal chain, c) S$_6$ ring molecule.

**Figure 6. Enthalpies (relative to the $R\bar{3}$ phase) of the phases of sulphur.** According to these calculations, at 0 K the $R\bar{3}$ structure is stable between 7 and 18 GPa. Figure 7. Structures of nitrogen

identified by USPEX at 100 GPa: a) *cubic gauche*, b) metastable polymeric form related to cubic gauche. *Cubic gauche* structure at 100 GPa: space group $I2_13$, $a$=3.515 Å, N=(0.0737;0.0737; 0.0737). Related metastable structure at 100 GPa: space group *C*2/*c*, $a$=4.943 Å, b=3.577 Å, $c$=4.874 Å, $\beta$=90.87°, N1=(0.3731;-0.0904;0.3016), N2=(0.3096;0.2646;0.3724).

**Figure 8. Structures of oxygen identified by USPEX at 25 GPa: a) stable chain structure, b) metastable structure with $O_4$ groups (higher in enthalpy by 7 meV/atom).** Stable structure at 25 GPa: space group *Cmcm*, $a$=6.514 Å, $b$=3.499 Å, $c$=2.871 Å, O=(0.0936;0.2926;1/4). Metastable structure: space group $P\bar{1}$, $a$=3.648 Å, $b$=4.489 Å, $c$=4.452 Å, $\alpha$=77.25°,$\beta$=70.68°,$\gamma$=71.53°, O1=(0.6840;0.4264;0.1965), O2=(0.3168;0.5022;0.2790), O3=(0.3186;0.1074;0.8032), O4=(0.6860;0.0315;0.7225).

**Figure 9. Structure of anatase ($TiO_2$) found by USPEX.**

**Figure 10. Stable high-pressure phases of $CaCO_3$ found by USPEX (see Ref. 72 for details).** a) Structure of post-aragonite (space group *Pmmn*): red spheres – Ca, blue – C, green – O atoms. b) calculated and experimental powder diffraction for samples quenched to 36 GPa: top – experiment, bottom – calculation (red peaks – post-aragonite, green – NaCl, blue – Pt), c) orthorhombic $C222_1$ structure stable above 137 GPa.

**Figure 11. Metastable phases of $CaCO_3$ found by USPEX in the range 50-80 GPa [72].** a) triclinic $P\bar{1}$, b) trigonal *R*3*m*, c) monoclinic $P2_1$. Between 50-80 GPa these metastable structures have enthalpies only 20-60 meV/atom above post-aragonite.

**Figure 12. Structures of ice ($H_2O$) at 1 atm found by USPEX.** a) Ice Ih and b) Ice Ic.

**Figure 13. Structure of urea, $(NH_2)_2CO$, found by USPEX.**

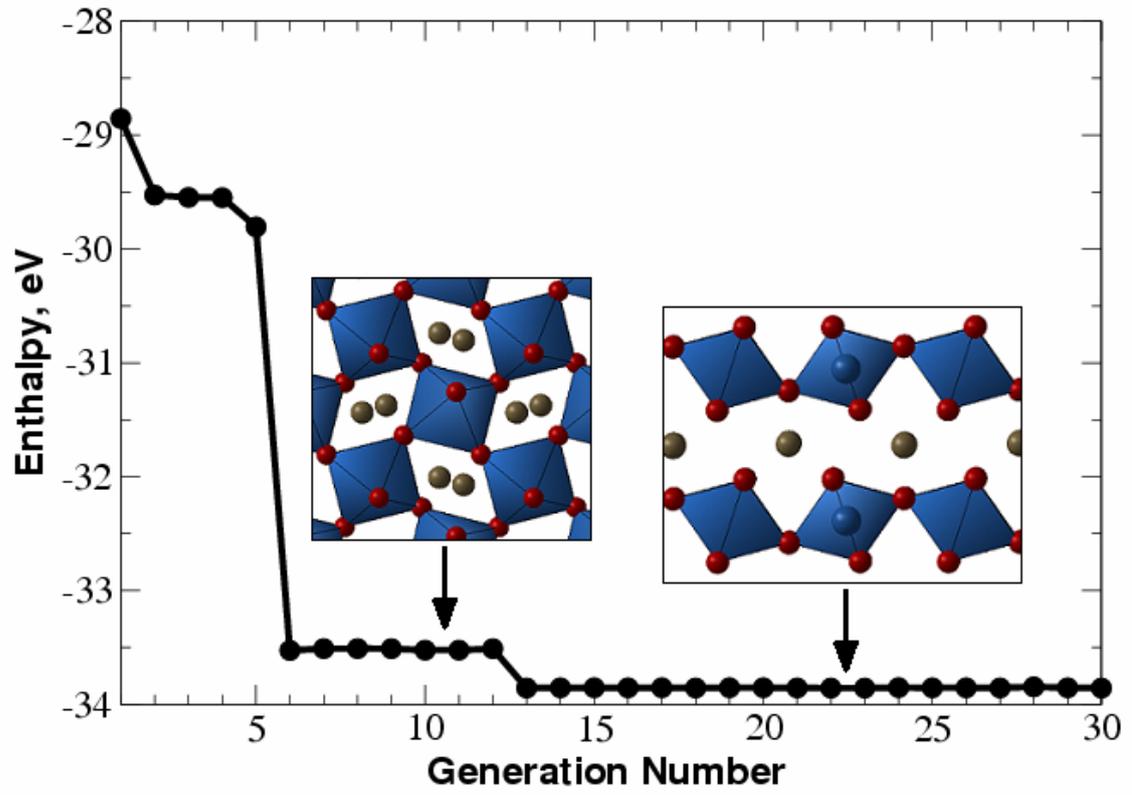

**Figure 1.**

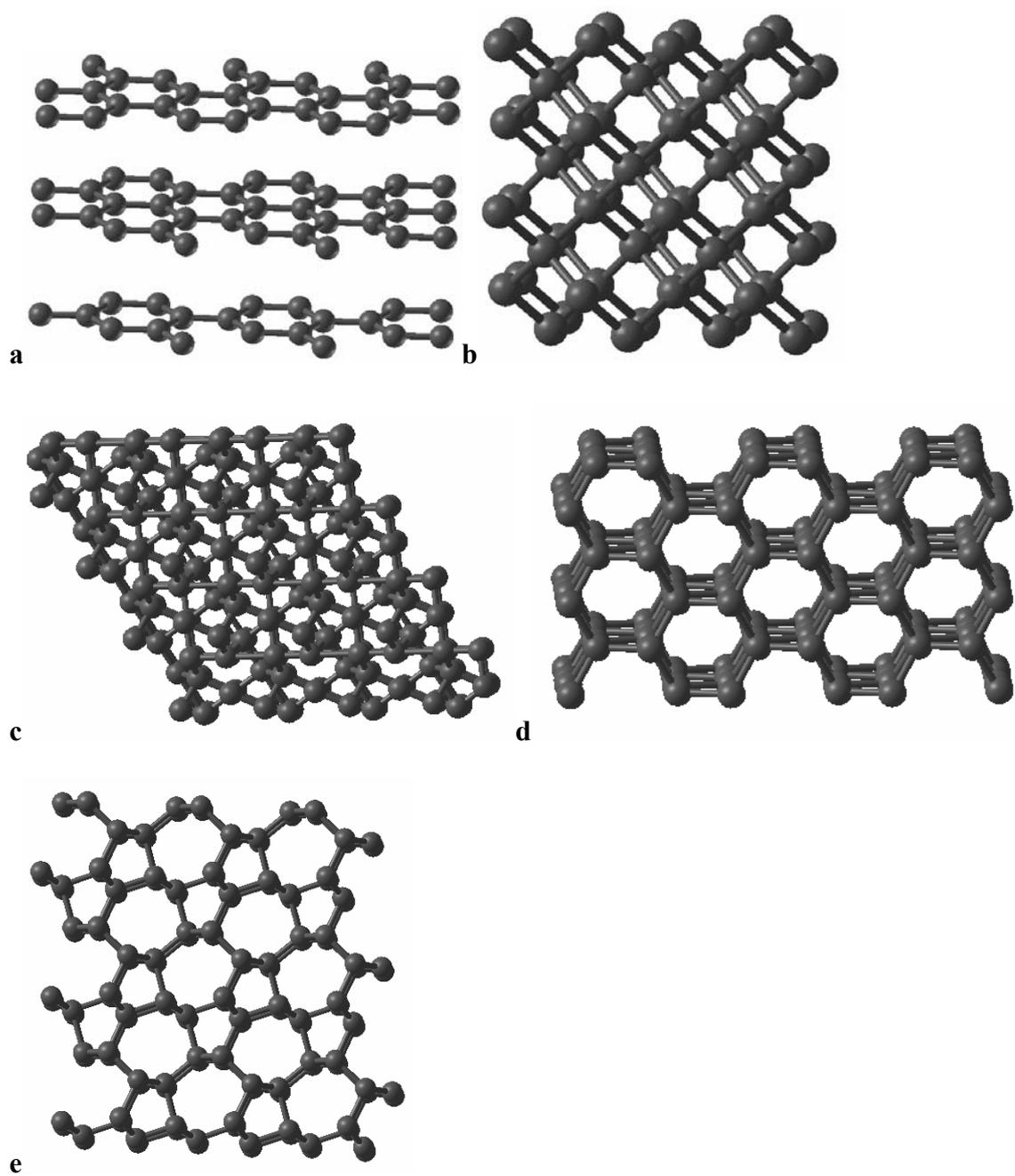

**Figure 2.**

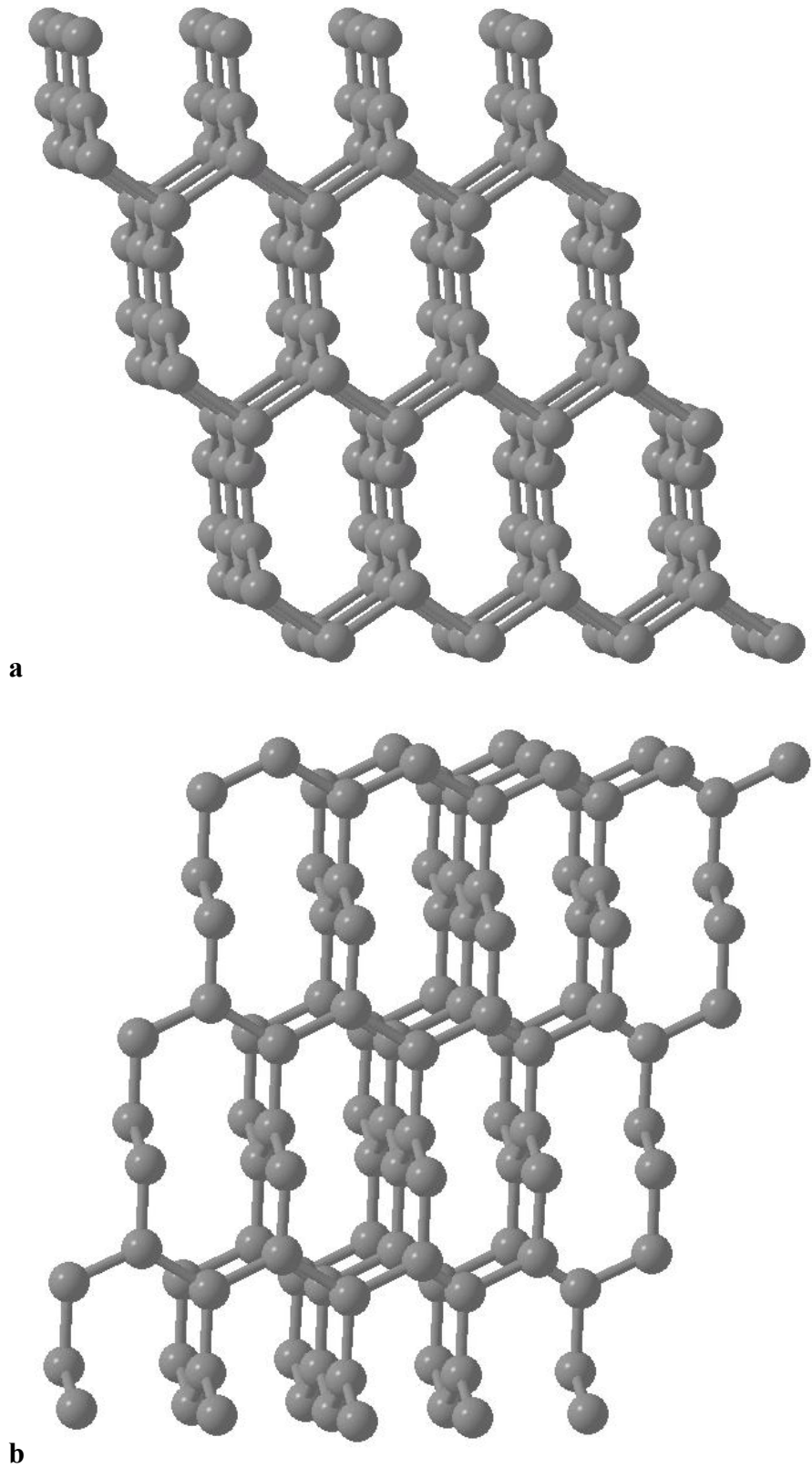

**a**

**b**

**Figure 3.**

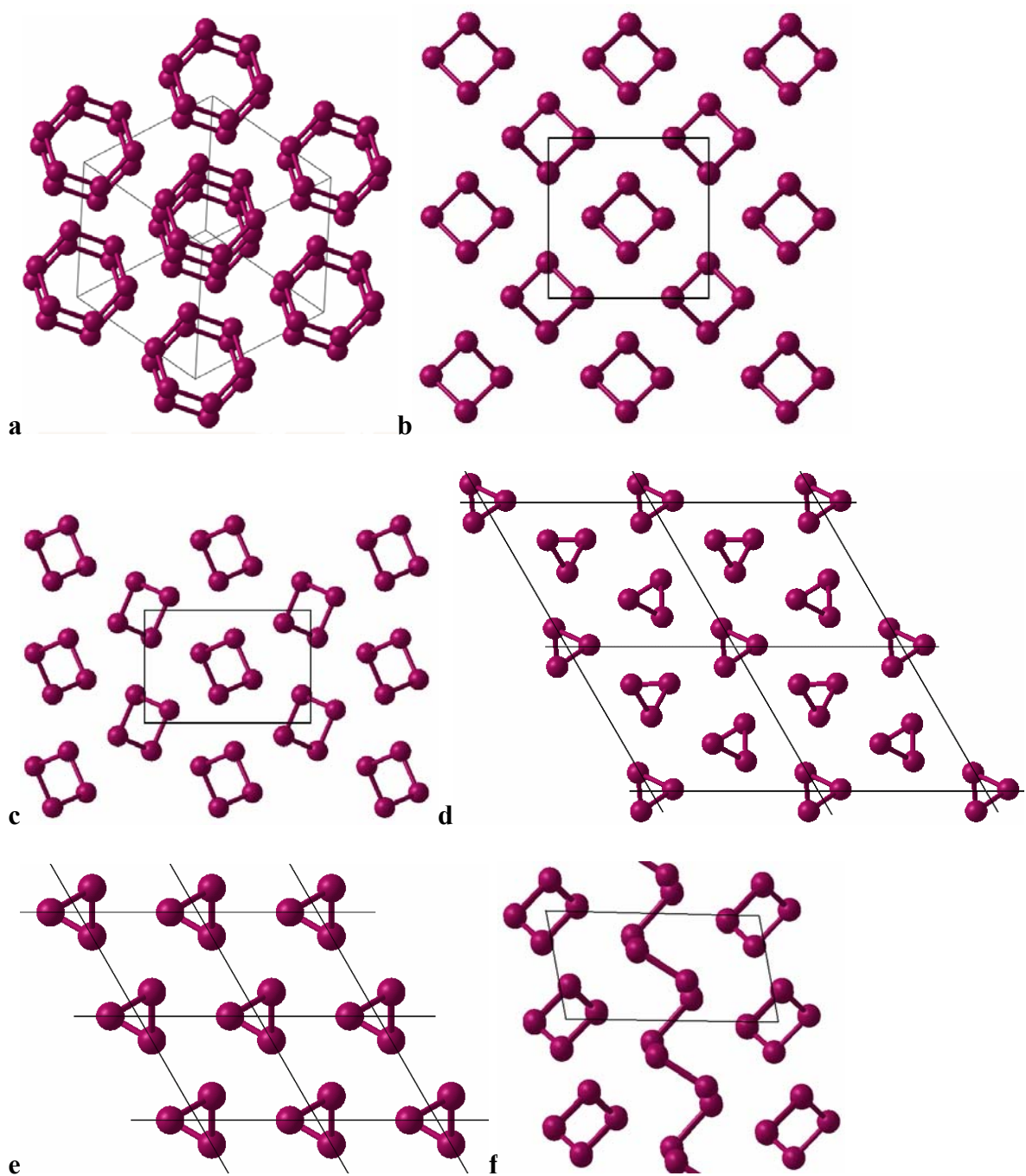

**Figure 4.**

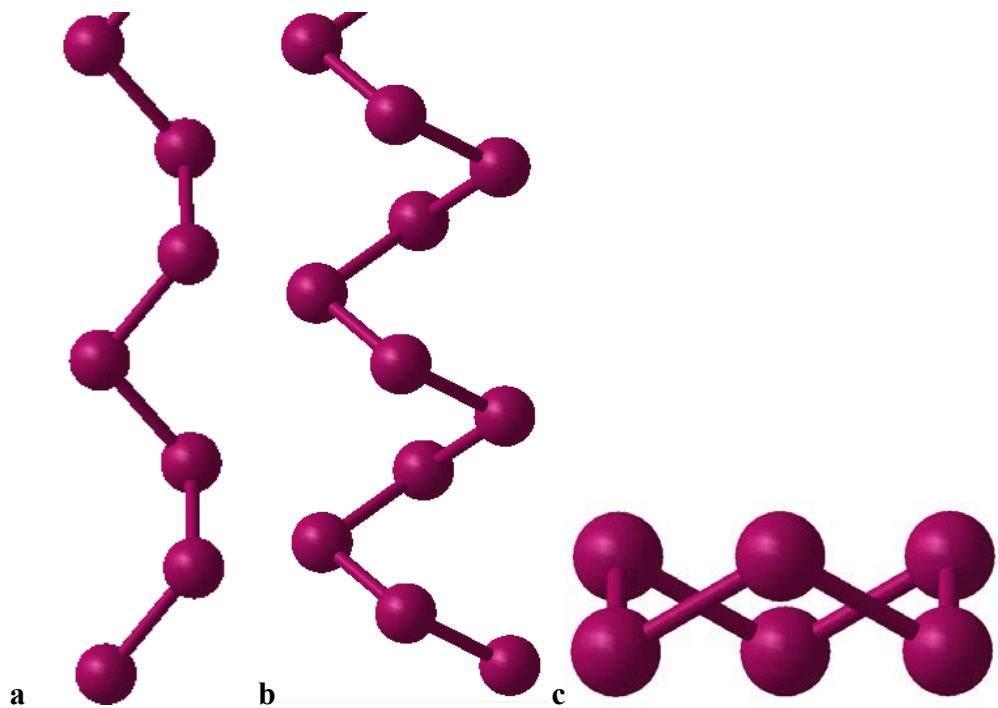

**Figure 5.**

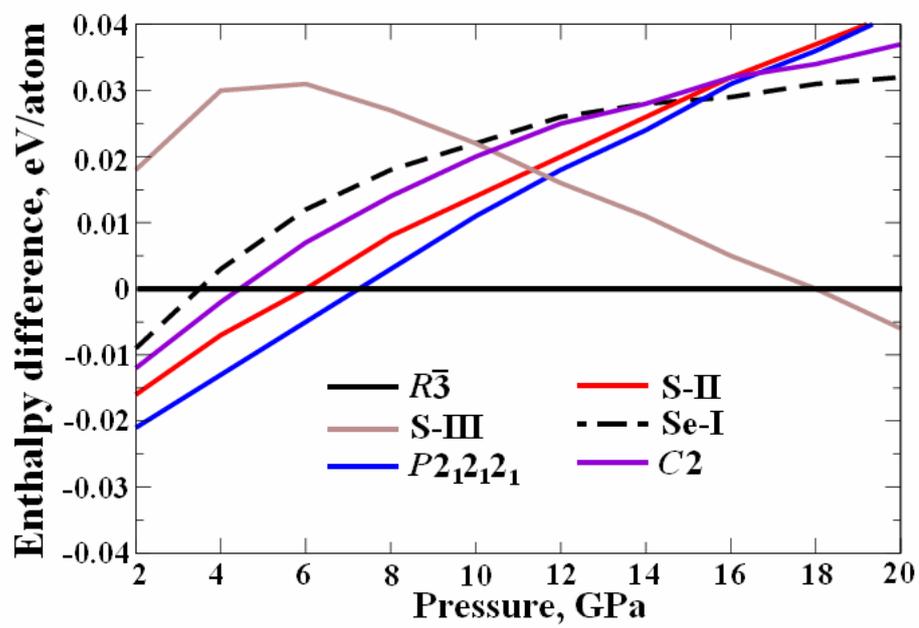

**Figure 6.**

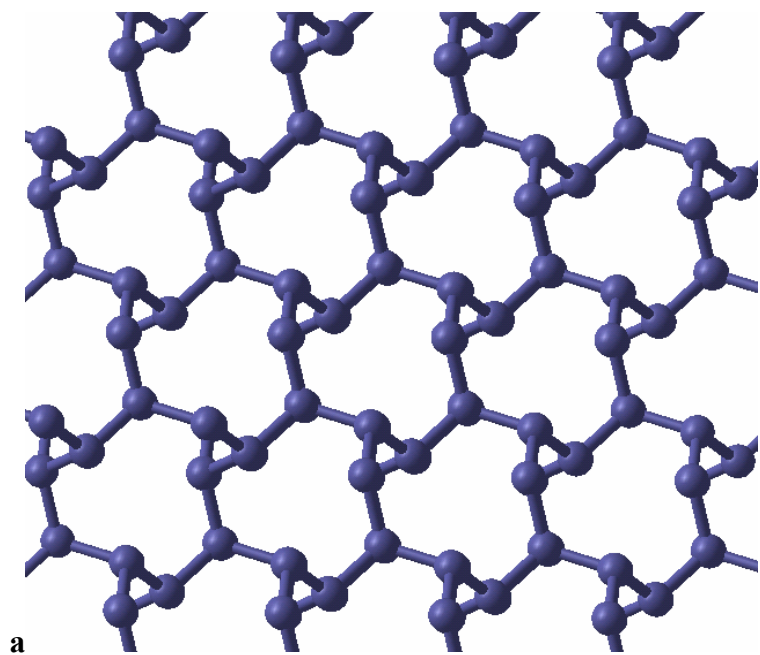

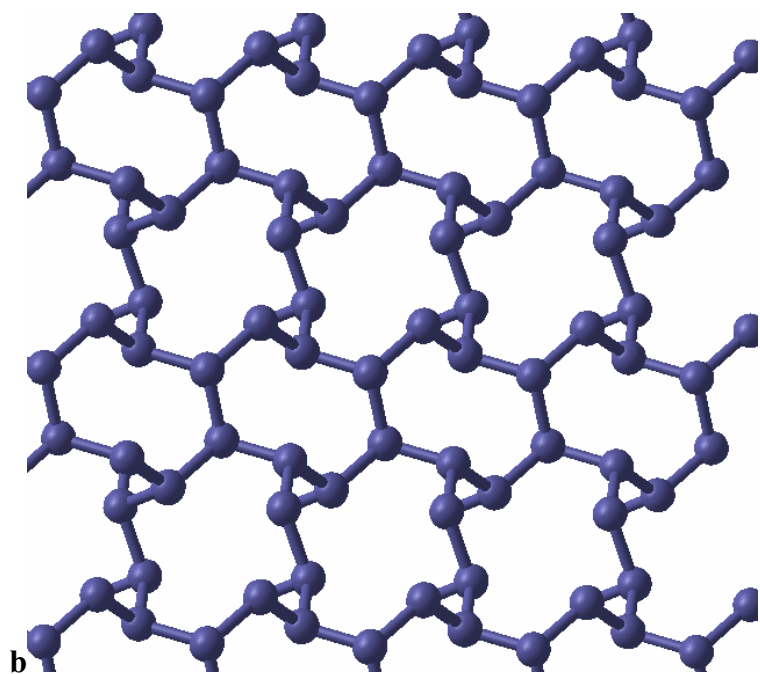

**Figure 7.**

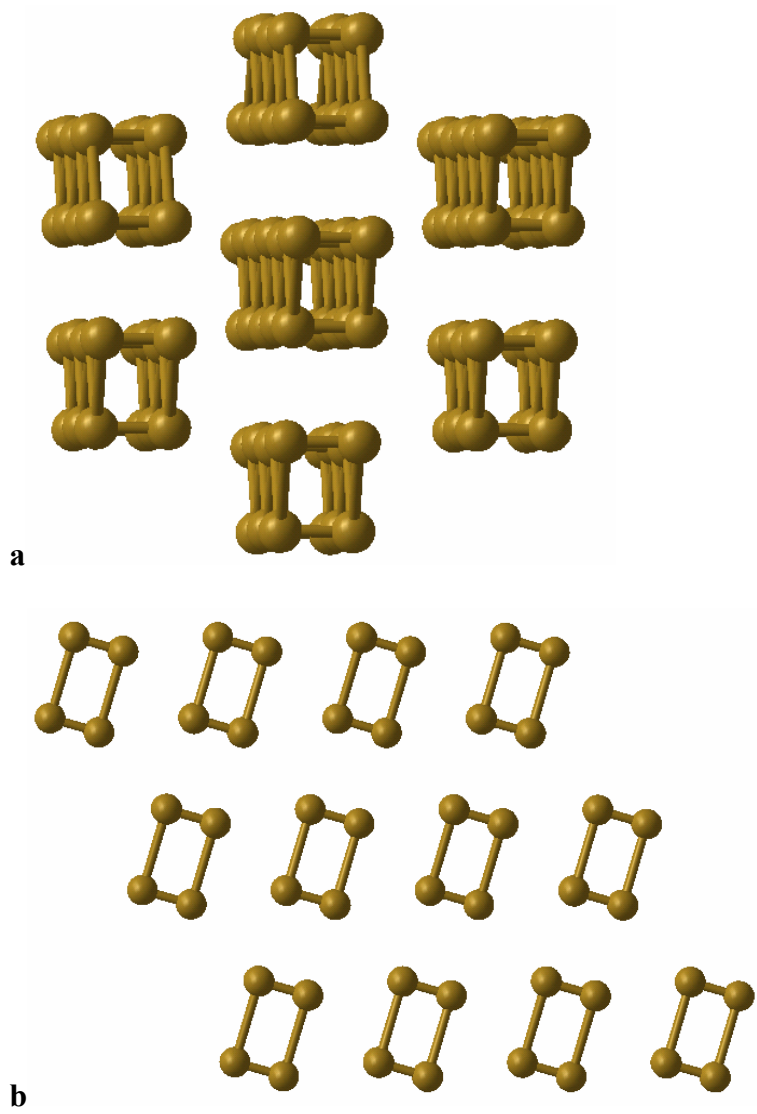

**a**

**b**

**Figure 8.**

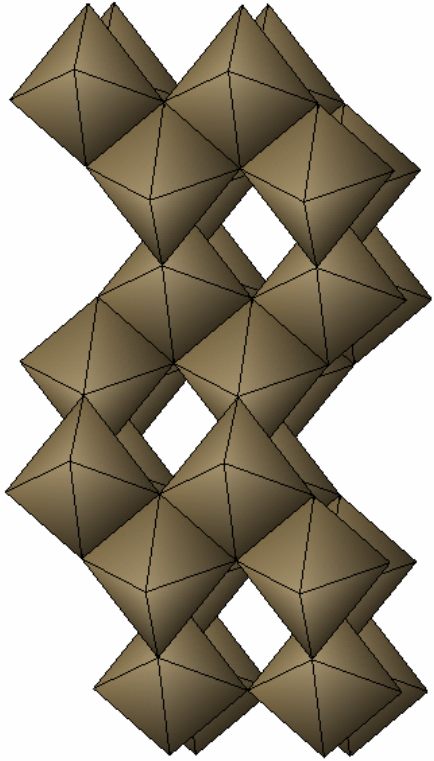

**Figure 9.**

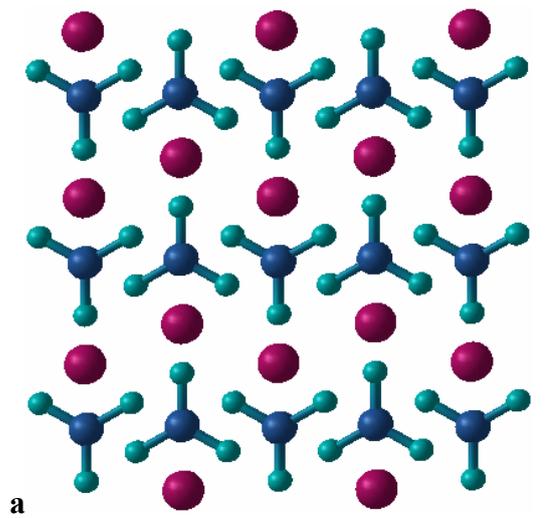

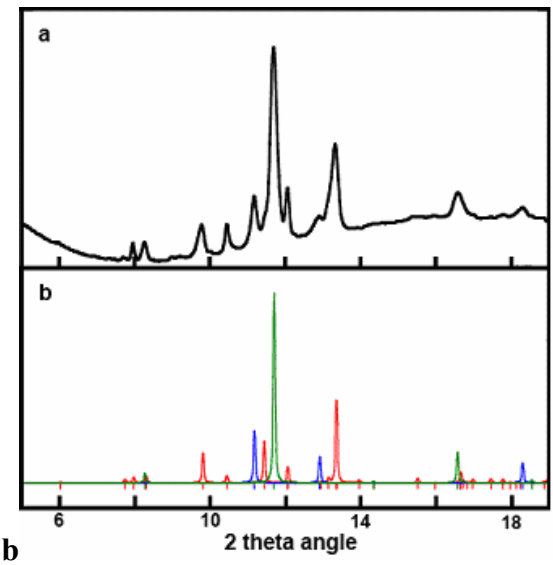

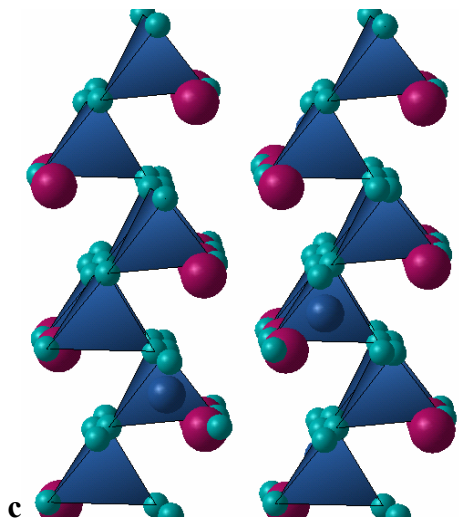

**Figure 10.**

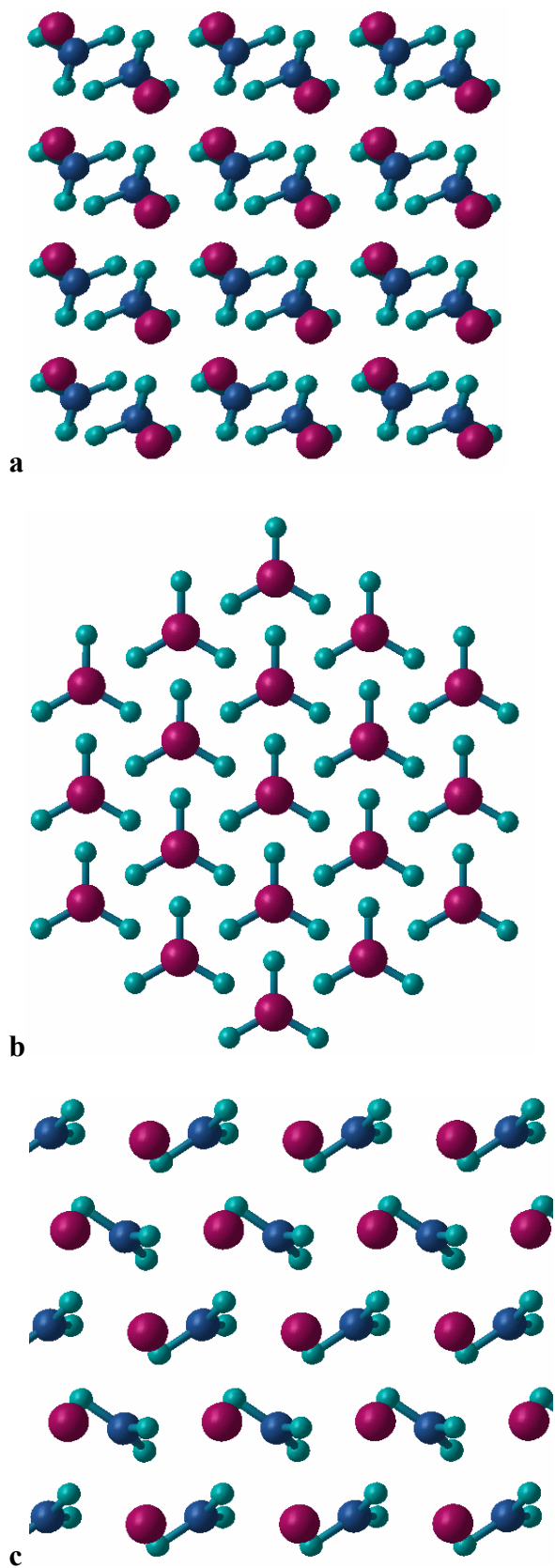

**Figure 11.**

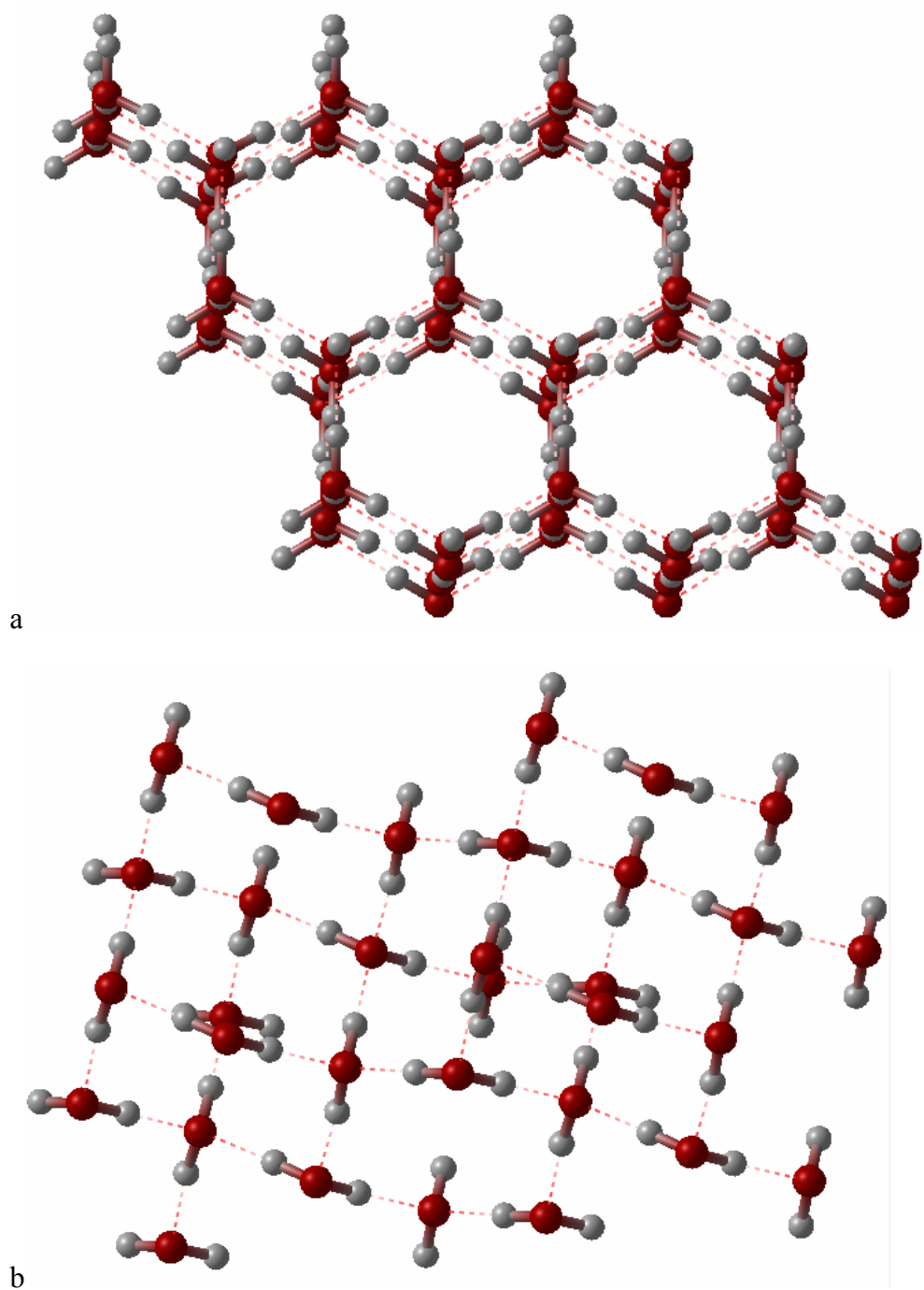

a

b

**Figure 12.**

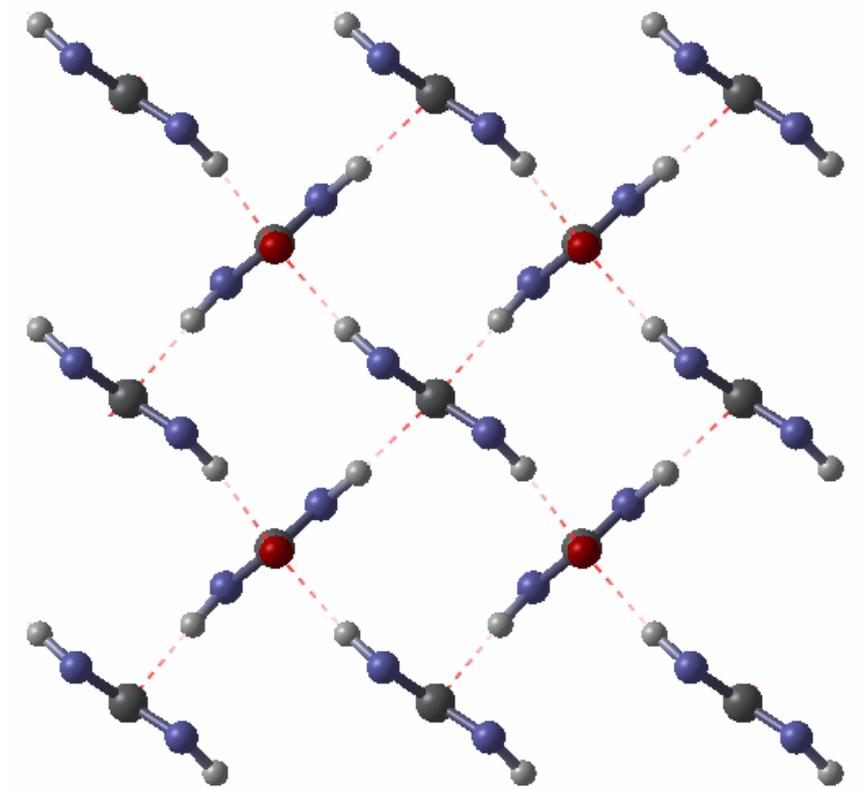

**Figure 13.**